\documentclass[]{spie}  %>>> use for US letter paper
\usepackage[dvips]{graphicx}

\def\lta{{\>\rlap{\raise2pt\hbox{$<$}}\lower3pt\hbox{$\sim$}\>}}
\def\gta{{\>\rlap{\raise2pt\hbox{$>$}}\lower3pt\hbox{$\sim$}\>}}

\title{Exploring the cosmic dark ages with the next generation of space and 
ground-based facilities}

\author{Massimo Stiavelli\supit{a} 
\skiplinehalf
\supit{a}Space Telescope Science Institute}

\authorinfo{Further author information: \\ E-mail: mstiavel@stsci.edu, Address: STScI, 
3700 San Martin Dr., Baltimore MD 21210
}
\begin{document} 
  \maketitle 

%%%%%%%%%%%%%%%%%%%%%%%%%%%%%%%%%%%%%%%%%%%%%%%%%%%%%%%%%%%%% 
\begin{abstract}
This paper reviews our current understanding of the process of re-ionization 
of the Universe, focusing especially on those models where re-ionization
is caused by UV radiation from massive stars. After reviewing the 
expected properties of stars at zero metallicity, I discuss the properties
of primordial HII regions and their observability.
\end{abstract}

%>>>> Include a list of keywords after the abstract 


%%%%%%%%%%%%%%%%%%%%%%%%%%%%%%%%%%%%%%%%%%%%%%%%%%%%%%%%%%%%%
\section{INTRODUCTION}
\label{sect:intro}  % \label{} allows reference to this section

The emergence of the first sources of light in the Universe and the
subsequent re-ionization of hydrogen marks the end of the "Dark Ages"
in cosmic history, a period characterized by the absence of discrete
sources of light. Despite its remote timeline, this epoch is
currently under intense theoretical investigation and will soon begin
to be probed observationally. The first reason to study this epoch is
the fact that this - together with the epoch of recombination - is 
the most accessible of the global phase
transitions undergone by the Universe after the Big Bang.

Two facts make the formation of structure in the Dark
Ages easier to study theoretically than similar processes occurring at
other epochs: {\it i)} the formation of the first structures is
directly linked to the growth of linear perturbations, and {\it ii)}
these objects have a known metallicity set by the end-product of the
primordial nucleosynthesis. Therefore, the second reason to study this epoch is
because it makes it possible to probe the
power spectrum of density fluctuations emerging from recombination at
scales smaller than accessible by current cosmic microwave background
experiments.

In a Universe where structures grow hierarchically, the first sources
of light act as seeds for the subsequent formation of larger
objects. Thus, the third reason to study this period is that by doing
so we may learn about processes relevant to the formation of the
nuclei of present day giant galaxies.

In Section \ref{sect:basics}, I will review our present view on the
basic processes leading to re-ionization. I refer the reader to the
recent review by Loeb and Barkana \cite{loebbarkana} for a more
detailed discussion. In Section \ref{sect:theory}, I will review in
more detail a few recent theoretical results relevant to the case of
re-ionization by stellar UV radiation. The present observational
status is summarized in Section \ref{sect:observations}. The
properties of primordial and low metallicity HII regions are discussed
in Sections \ref{sect:primordial} and \ref{sect:lowmetal},
respectively. Finally, in Section \ref{sect:future}, I address the
observability of these objects by NGST and future large ground-based
telescopes.

\section{THE BASIC PROCESSES LEADING TO RE-IONIZATION}
\label{sect:basics}

In the standard Cold Dark Matter (CDM) cosmology galaxies are
assembled through hierarchical merging of building blocks with smaller
mass. The first such building blocks, with M$\gta 10^4$
M$_\odot$
\cite{CouchmanRees1986,HaimanLoeb1997,OstrikerGnedin1996,HaimanThoulLoeb96,Abeletal98,Abeletal00},
form in various CDM models at z$\gta$15. 

While we do not know whether the first sources of light are powered by
nuclear energy from fusion reactions or by gravitational
accretion\cite{HaimanLoeb1999}, it is often believed that population III
stars are responsible for the re-ionization of
hydrogen\cite{MadauShull} at z$\simeq$6-20
\cite{HaimanLoeb1999,GnedinOstriker97,ChiuOstriker00} while the harder UV
spectrum of AGNs is responsible for the re-ionization of
helium\cite{Jacobsen} at lower redshift.

%-------------
   \begin{figure}
   \begin{center}
   \begin{tabular}{c}
   \includegraphics[height=10cm]{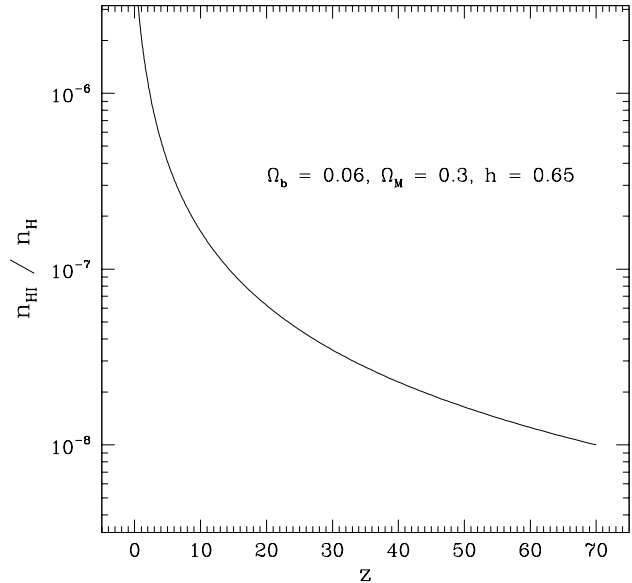}
   \end{tabular}
   \end{center}
   \caption[example] 
%>>>> use \label inside caption to get Fig. number with \ref{}
   { \label{fig:GunnPeterson} The fraction of neutral hydrogen needed 
to produce an average $\tau=1$ shortwards of Lyman $\alpha$ is shown
as a function of redshift. The plot is valid for a
uniform IGM. Only a small fraction of neutral hydrogen along the line
of sight would be sufficient to produce significant opacity.}
\end{figure}
%-------------

The following simple calculation shows that nuclear processing of only
a very small fraction of baryons would be sufficient to re-ionize the
Universe\cite{loebbarkana}. Fusion of hydrogen releases 7 MeV per
proton but only 13.6 eV are needed to ionized hydrogen. Thus, a
fraction 2$\times 10^{-6}$ of hydrogen undergoing fusion is
energetically sufficient to re-ionize all hydrogen. In practice, the
minimum fraction will be larger by a factor ten to one hundred because
not all energy is released in the form of ionizing photons, not all
ionizing photons successfully escape from the sources into the IGM, 
and recombinations increase the
required number of ionizing photons. All these factors depend on the
details of the process and have large uncertainties. By assuming that
the minimum fraction is 30 times the value given above, i.e., 6$\times
10^{-5}$ and a yield of $\sim$0.3 $M_\odot$ of metals produced per
$M_\odot$ of hydrogen processed in a massive
star\cite{HegerWoosley2002,Ohetal2001}, one can estimate that the
minimum mean metallicity of the Universe at re-ionization was $\sim
10^{-3} Z_\odot$.

If the power source for re-ionization is not nuclear fusion but rather
gravitational accretion onto massive black holes, a smaller fraction
of material needs to be processed thanks to the higher efficiency of
gravitational accretion. Clearly this scenario does not place any
constraint on the metallicity of the Universe at re-ionization and
leaves unanswered the question of the origin of the black holes (primordial
or stellar). Even if re-ionization is caused by stellar UV radiation,
it is natural to expect that some fraction of these stars will leave
black holes as remnants\cite{MadauRees2001}. Thus, in both scenarios,
we expect that some seed black holes will be present at the end of
re-ionization, with interesting implications on the formation of AGNs and
galaxies\cite{SilkRees,StiavelliBH}.

Even though we often refer to re-ionization as if it was a sudden transition,
the time elapsed between the epochs when 10
\% and 90 \% of hydrogen was re-ionized can last a significant fraction
of the age of the Universe at re-ionization: thus, the dark ages may end with
an extended twilight. 
Inhomogeneities along the line of sight may create a dispersion in
optical depth shortwards of Lyman $\alpha$.  Moreover, only a very low
residual fraction of neutral hydrogen is needed to produce a
Gunn-Peterson trough in the spectra of high redshift quasars.  To
illustrate this I give in Figure \ref{fig:GunnPeterson}, as a function
of redshift, the fraction of neutral hydrogen required to have a
Gunn-Peterson optical depth of one in a uniform
IGM\cite{GunnPeterson,Beckeretal2001}. It should be noted that in
analogy to the proximity effect in QSOs\cite{PalleMoller} the
opacity near Lyman $\alpha$ would be modified
in the neighborhood of ionizing sources\cite{MiraldaRees94}.
In order to obviate the problem of the extreme sensitivity of the Gunn-Peterson
trough to even a very small residual fraction of neutral hydrogen, methods
are being considered that rely on the study of the microwave background
polarization\cite{Kaplinghatetal}.

In the following I will assume that re-ionization is due to massive
stars and explore the theoretical and observational consequences of
this assumption.

\section{RE-IONIZATION BY STELLAR UV RADIATION : SOME THEORETICAL 
CONSIDERATIONS}
\label{sect:theory}

If re-ionization of hydrogen is caused by massive stars, the details of
the process and its consequences for the metal enrichment of the
Universe depend on the properties of these objects and on how metals
and UV radiation escape into the IGM. Below, I will briefly review
present ideas on some of these issues.

%-------------
   \begin{figure}
   \begin{center}
   \begin{tabular}{c}
   \includegraphics[height=10cm]{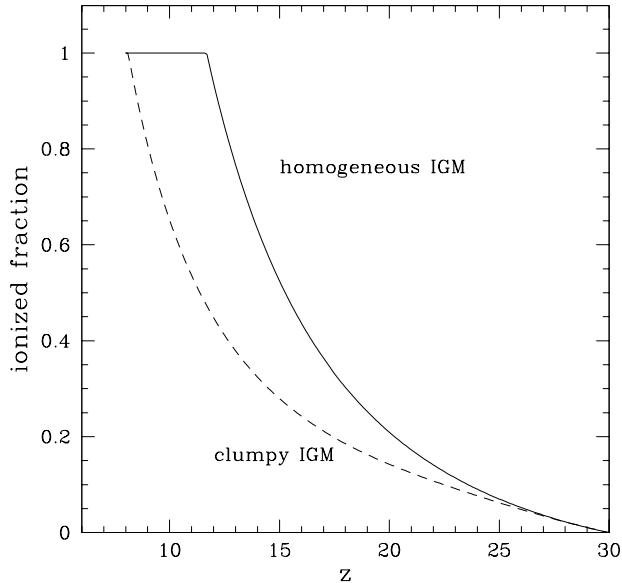}
   \end{tabular}
   \end{center}
   \caption[example] 
%>>>> use \label inside caption to get Fig. number with \ref{}
   { \label{fig:madaureion} The fraction of ionized
hydrogen as a function of
redshift for a Madau et al \cite{Madauetal98} star formation rate extrapolated to z=30. 
The solid line applies to the case of a uniform IGM while the dashed line 
represents a clumpy IGM with clumpiness parameter C=10 (see text).}
\end{figure}
%-------------

\subsection{Which stars?}

The standard picture is that, at zero metallicity, cooling is
dominated by the less effective H$_2$ cooling and leads to the
formation of very massive objects, with masses exceeding 100
$M_\odot$\cite{BrommCoppiLarson99,BrommCoppiLarson02}. The spectral
energy distribution (SED) of these massive stars resembles 
that of a black body with an effective temperature around $10^5$
K\cite{BrommKudritzkiLoeb01}. Due to their high temperatures, these stars
are very effective in ionizing hydrogen and helium.

It should be noted that, even at lower mass, zero-metallicity stars are
expected to be much hotter than their solar metallicity
analogues\cite{TumlisonShull00}. By assuming that only massive stars
are present, one finds that roughly $\sim 10^{-4}$ of baryons need
to be incorporated into stars in order to re-ionize
hydrogen\cite{BrommKudritzkiLoeb01}, in agreement with the estimates of
Section \ref{sect:basics}.

The second generation of stars forming out of pre-enriched material
will probably have different properties since molecular hydrogen can
be dissociated by UV radiation of lower energy ($\sim$11 eV) than that
required to ionize atomic
hydrogen\cite{StecherWilliams67,HaimanReesLoeb97}. This implies that
molecular hydrogen cannot be shielded by atomic hydrogen. However,
the presence of some metals in the ISM opens up the possibility of having
some shielding by dust. Various
scenarios have been predicted for the second generation objects. If
the metallicity is high enough ($>5\times 10^{-4} Z_\odot$), cooling by
metal lines is viable and one is likely to produce stars of lower
mass\cite{BrommFerraraetal01}. On the other hand, if the metallicity
is lower, halos with virial temperature greater than $10^4$ K can cool
via neutral atomic lines. Such systems may experience a build-up of
H$_2$ thanks to self-shielding and continue the formation of very
massive stars\cite{OhHaiman02}.

%-------------
   \begin{figure}
   \begin{center}
   \begin{tabular}{c}
   \includegraphics[height=10cm]{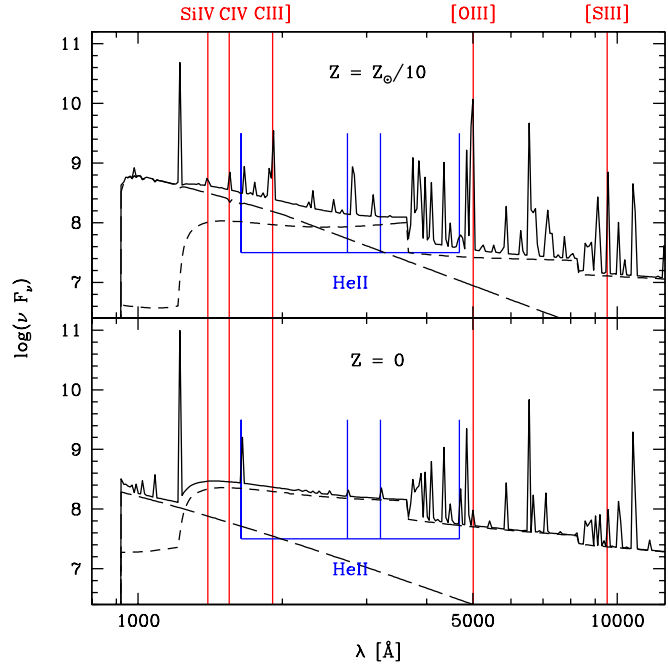}
   \end{tabular}
   \end{center}
   \caption[example] 
%>>>> use \label inside caption to get Fig. number with \ref{}
   { \label{fig:zerometalHII} 
The synthetic spectrum of a zero-metallicity HII region (bottom panel)
is compared to that of an HII region with Z = 0.1 Z$_\odot$. The long
dashed lines represent the stellar continuum while the short dashed
lines represent the nebular continuum. Note how the latter dominates
the continuum in the zero-metallicity case. }
   \end{figure} 
%-------------

\subsection{Which Initial Mass Function?}

From the previous discussion it appears that in the zero-metallicity
case one should expect a very top heavy Initial Mass Function
(IMF). It is less clear whether the second generation of stars is also
top-heavy or characterized by a more normal IMF. In practice, it is wise to
consider both possibilities.

\subsection{What about re-combinations? }

The number of UV photons needed to re-ionize the Universe depends on
the rate of hydrogen recombination. Recombinations are not very important if
the medium is homogeneous.  However, inhomogeneities increase locally
the density and therefore can increase
dramatically the recombination rate which is proportional to the square of
the local density. The basic features of this effect can be captured
by a simple model\cite{ShapiroGiroux87}. In Figure
\ref{fig:madaureion} I have considered the star formation rate (SFR)
given by Madau et al. \cite{Madauetal98} extrapolated to
z=30. Assuming this SFR and a uniform IGM, the Universe would reionize
at $z_r=11.7$ and recombinations would increase the required number of
ionizing photons by only 28 \% relative to the case without recombination.  
The same star formation history in a Universe with a clumpiness factor\cite{loebbarkana} 
$C\equiv <n_H^2>/<n_H>^2$ = 10 would
re-ionize hydrogen at $z_r = 8.1$, with on average 3.4 photons required
to produce a single re-ionization. These results do not depend
critically on the specific starting redshift.

It is possible to calculate in more detail the amount of
inhomogeneity\cite{Miraldaetal}.  As an example, it has been argued
that if re-ionization occurs at $z_r < 20$ a significant fraction of
mini-halos ($\sim 10^4 M_\odot$) will be able to collapse and, in the
process of being photo-evaporated, act as significant sinks of UV
radiation\cite{HaimanAbelMadau01,BarkanaLoeb02}. The net effect of an increased
recombination rate will be a higher mean metallicity of the Universe
at re-ionization, perhaps by a factor 3.

%-------------
   \begin{figure}
   \begin{center}
   \begin{tabular}{c}
   \includegraphics[height=10cm]{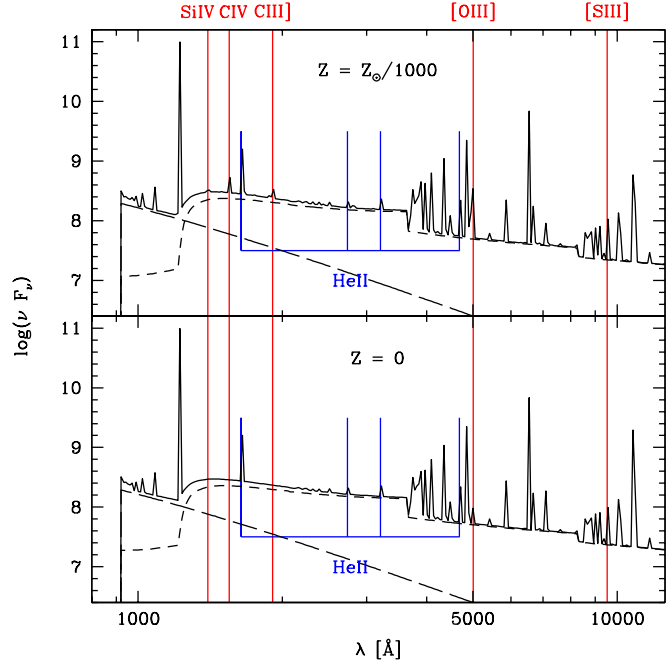}
   \end{tabular}
   \end{center}
   \caption[example] 
%>>>> use \label inside caption to get Fig. number with \ref{}
   { \label{fig:lowmetalHII} 
The synthetic spectrum of a zero-metallicity HII region (bottom panel)
is compared to that of an HII region with Z = $10^{-3} Z_\odot$. The
long dashed lines represent the stellar continuum while the short
dashed lines represent the nebular continuum. Note how the latter
dominates the continuum in both cases. }
   \end{figure} 
%-------------

\subsection{What about the escape of UV radiation?}

If the first generation stars forms within more massive objects only a
fraction of their UV radiation may be able to escape. For local disk
galaxies the escape fraction is probably of only $\lta 10$\%
\cite{Leithereretal95,DoveShull94,DoveShullFerrara00}. The estimate
for high redshift galaxies could be as high as 50 \%
\cite{Steideletal01} but could be lower if UV-grey dust is present in
the sample\cite{Huietal02}. If only a fraction of UV radiation escapes
from the first objects this implies once again that the mean metallicity
of the Universe will be higher if stars are the relevant UV sources.

\subsection{How is the Inter Galactic Medium enriched? }

When considering the mean metallicity of the Universe at
re-ionization, we have been referring to the total amount of metals
produced. This is in general very different from the metallicity
actually measured in the Inter-Galactic Medium (IGM). If population
III stars are formed in halos of sufficiently low mass they can enrich
the IGM by SN-driven winds\cite{MadauFerraraRees01,MoriFerraraMadau02}.  
When a halo undergoes a SN-driven outflow, the
ejection of metals can be very effective. However, it is not clear
how effective this process is when averaged over all halos.

Note that if the escape fractions of UV radiation and metals are
comparable, we should expect a mean metallicity of the IGM at
re-ionization similar to the minimum mean metallicity of the Universe,
$Z \simeq 10^{-3} Z_\odot$ (see Sect \ref{sect:basics}). This is
broadly consistent with the metallicity of damped Lyman $\alpha$
systems which is higher than $10^{-3} Z_\odot$ for
$z<4$\cite{Prochaskaetal01}.

%-------------
   \begin{figure}
   \begin{center}
   \begin{tabular}{c}
   \includegraphics[height=10cm]{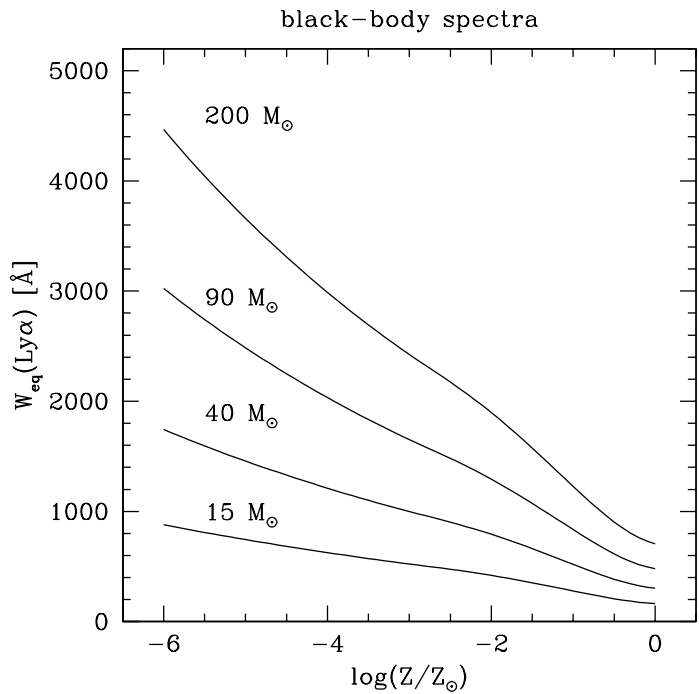}
   \end{tabular}
   \end{center}
   \caption[example] 
%>>>> use \label inside caption to get Fig. number with \ref{}
   { \label{fig:LyalphaEW} 
Lyman $\alpha$ equivalent widths for HII regions ionized by stars with
a range of masses and metallicities. The results obtained for black bodies
or stellar atmospheres are very similar.}
   \end{figure} 
%-------------

\subsection{Remnants and signature of a population III}
	
A fraction of the supermassive stars produced at zero metallicity may
leave massive black holes as remnants\cite{HegerWoosley2002}. It has
been argued that under a reasonable assumption the mass density of these
black holes may be comparable to that of the supermassive black holes observed
locally in the cores of giant galaxies\cite{MadauRees2001}. This opens up
the interesting possibility that the supermassive black holes are formed
by merging of the population III remnants. 

Population III objects are characterized by nucleosynthetic patterns
different from those produced by ordinary
stars\cite{HegerWoosley2002,Ohetal2001}.  In principle, such patterns
could be identified by studying low metallicity objects at low
redshift but possible contamination throughout cosmic history makes this 
measurements difficult to interpret. Thus, it may be better to look for 
population III signatures in the Lyman $\alpha$ forest absorbers at 
high redshift.

%-------------
   \begin{figure}
   \begin{center}
   \begin{tabular}{c}
   \includegraphics[height=10cm]{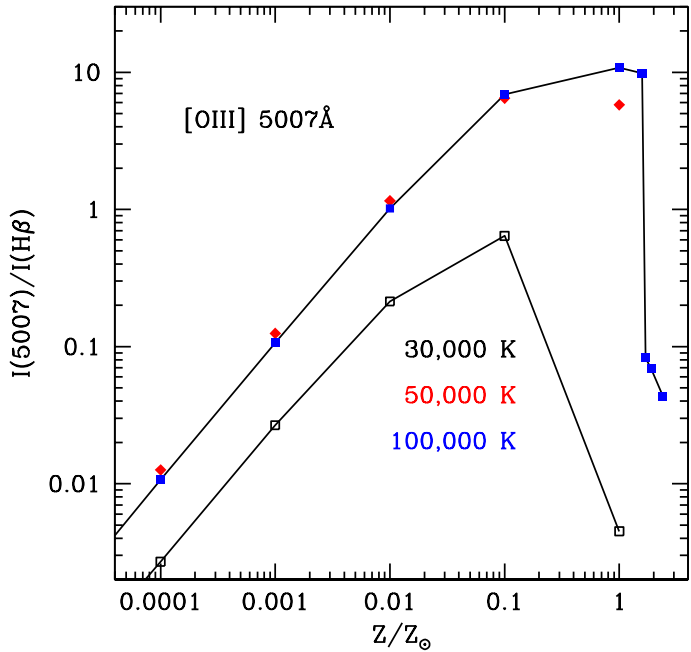}
   \end{tabular}
   \end{center}
   \caption[example] 
%>>>> use \label inside caption to get Fig. number with \ref{}
   { \label{fig:OIIIcalibration} 
The ratio [OIII]$\lambda 5007$ / H$\beta$ is plotted as a function of
metallicity for three different effective temperatures: 30,000K (open squares and
bottom line), 50,000K (solid diamonds), and 100,000K (solid squares and top
line).}
   \end{figure} 
%-------------

\section{PRESENT OBSERVATIONAL STATUS}
\label{sect:observations}

The most direct observational evidence of re-ionization is the
detection of a Gunn-Peterson trough\cite{GunnPeterson} in the
spectrum of high redshift quasars. Neutral hydrogen clouds along the
line of sight (the Lyman $\alpha$ forest) produce increasing
absorption as the redshift increases. Still, even at z$\simeq$5 some
signal is detected below Lyman $\alpha$, suggesting that re-ionization
occurs at higher redshifts.

Recently, the Sloan Digital Sky Survey has began detecting large
numbers of high redshift quasars, including a few around z$\simeq$6
\cite{sloanqsoz6}. QSO SDSSp J103027.10 0552455.0 at z=6.28 shows a
drop in continuum flux below Lyman $\alpha$
by a factor 150. Other
QSOs at slighly lower redshift show a much smaller continuum drop.
%(see Figure \ref{fig:qsoz6}. 
Thus, it has been suggested that a
Gunn-Peterson trough has been detected in this
object\cite{Beckeretal2001,Fanetal2002}.

On the basis of the qualitative arguments given in Sect \ref{sect:basics}, 
it is clear that it is hard to make a statement regarding re-ionization
on the basis of a single object. However, this detection opens up the
possibility that re-ionization occurs at a relatively low redshift,
thus making it more easily accessible to observations.

Unlike hydrogen re-ionization, that of helium has
been already firmly identified from the detection of a
Gunn-Peterson trough \cite{Jacobsen,Davidsenetal,Heapetal}. 
Due to the higher photon energy
required to re-ionize helium, it is often assumed that it is caused by
the harder UV radiations from AGN.

\section{PRIMORDIAL HII REGIONS}
\label{sect:primordial}

Two direct consequences of the high effective temperature of
zero-metallicity stars are their effectiveness in ionizing hydrogen
(and helium) and their low optical-to-UV fluxes.  Both tend to make
the direct detection of the stellar continuum much harder than the
detection of the associated HII region.

In Panagia et al.\cite{Panagiaetal02} we report on our calculations
using Cloudy90\cite{Ferlandetal98} of the properties of these HII
regions (see also Figure \ref{fig:zerometalHII}.) We find that the
electron temperatures is in excess of 20,000 K and that 45 \% of the
total luminosity is emitted by the HII region in the Lyman $\alpha$ line, 
resulting in a
Lyman $\alpha$ equivalent width (EW) of 3000 \AA
\cite{BrommKudritzkiLoeb01}. The helium lines are also rather strong,
with the intensity of HeII $\lambda$1640 comparable to that of
H$\beta$\cite{Panagiaetal02,Tumlinsonetal01}

An interesting feature of these models is that the emission longward
of Lyman $\alpha$ is dominated by a strong two-photon nebular
continuum. The H$\alpha$/H$\beta$ ratio for these models is 3.2. Both
the red continuum and the high H$\alpha$/H$\beta$ ratio could be
naively (and incorrectly) interpreted as a consequence of dust
extinction even though no dust is present in these systems.

From the observational point of view one will generally be unable to
measure a zero-metallicity but will usually be able to place an upper
limit on it. When would such an upper limit be indicative that one is
dealing with a population III object? On the basis of the arguments in
the previous sections it is reasonable to argue that a metallicity
Z$\simeq10^{-3}Z_\odot$ can be used as a dividing line between the pre
and post re-ionization Universe. A similar value is obtained by
considering that the first supernova (SN) going off in a primordial
cloud with a mass of 10$^6$ M$_\odot$ will pollute it to a metallicity 
of $\sim 0.5 \times 10^{-3}Z_\odot$\cite{Panagiaetal02}. Computing the
number of SNae required to unbind the ISM in a primordial cloud
provides an independent - and higher - value for the metallicity, 
Z$\simeq10^{-2} Z_\odot$\cite{FallRees85}.
Thus, any object with a
metallicity higher than $\sim 10^{-3} Z_\odot$ is not a true first
generation object.

%-------------
   \begin{figure}
   \begin{center}
   \begin{tabular}{c}
   \includegraphics[height=10cm]{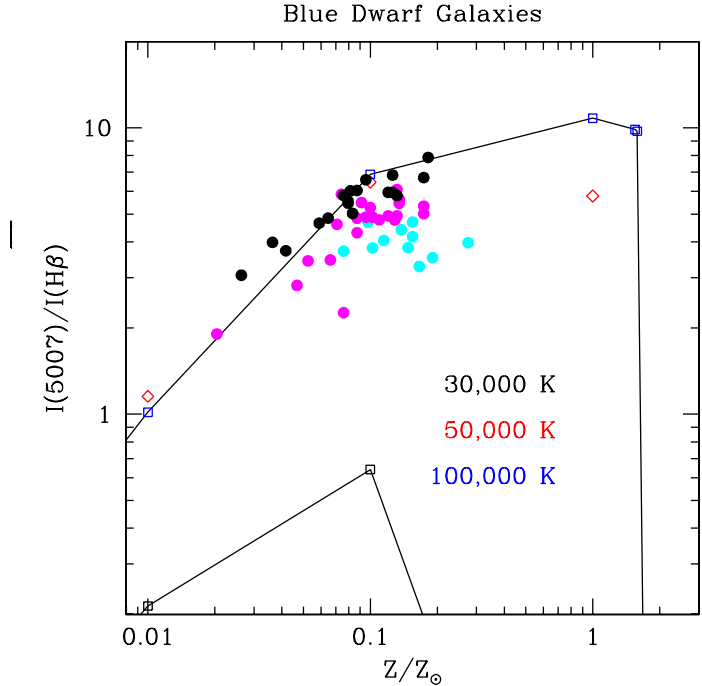}
   \end{tabular}
   \end{center}
   \caption[example] 
%>>>> use \label inside caption to get Fig. number with \ref{}
   { \label{fig:dwarfs} 
Blue dwarf galaxies can be used to verify the theoretical calibration
of [OIII]/H$\beta$ as a metallicity indicator. Models derived for
three different stellar masses are considered : 30,000K (open squares
and bottom line), 50,000K (solid diamonds), and 100,000K (solid
squares and top line). The black solid circles refer to dwarf galaxies
with [OIII]$\lambda$5007/[OII]$\lambda$3727 $>$ 5, the dark grey
circles to dwarf galaxies with [OIII]/[OII] in the range 2.5-5, and
the light grey circles to galaxies with [OIII]/[OII]$<$2.5. }
   \end{figure} 
%-------------

\section{LOW METALLICITY HII REGIONS}
\label{sect:lowmetal}

In order to distinguish a zero-metallicity HII region from one with
very low, but non-zero, metallicity we need to compute the properties
of the latter objects. We have carried out such an analysis for
metallicity down to $10^{-6} Z_\odot$\cite{Panagiaetal02} using a
combination of 
Cloudy90\cite{Ferlandetal98} and analytical calculations. I
n Figure \ref{fig:lowmetalHII} the
synthetic spectrum of an HII region with metallicity $10^{-3} Z_\odot$
is compared to that of an object with zero metallicity. The two are
very similar except for a few weak metal lines with the
strongest one being the [OIII]$\lambda$5007 line.  In Figure
\ref{fig:LyalphaEW} we show the Lyman $\alpha$ EWs for individual
HII regions
ionized by stars with different masses and metallicities. Values of
EW in excess of 1,000 are possible already for objects with
metallicity $\sim 10^{-3} Z_\odot$. This is particularly interesting
given that Lyman $\alpha$ emitters with large EW have been identified
at z=5.6\cite{RhoadsMalhotraz5} and possibly at even higher redshift
\cite{Huetal}.

Even though the metal lines at these metallicities are weak, some of them
can be used as metallicity tracers. In Figure
\ref{fig:OIIIcalibration} the ratio of the intensity of [OIII]$\lambda
5007$ to H$\beta$ is plotted for a range of stellar temperatures and
metallicities. It is immediately apparent that for $Z < 10^{-1}
Z_\odot$ this line ratio traces metallicity linearly (for each
individual star.) Our reference value $Z = 10^{-3}$ corresponds to a
ratio [OIII]/H$\beta$ = 0.1. The weak dependence on mass ensures
that this ratio remains an excellent indicator of metallicity also when one
considers the integrated signal from a population with a range of
stellar masses. The validity of these conclusions is confirmed by
comparing the model predictions to 
the properties of blue dwarf galaxies (see Figure
\ref{fig:dwarfs}).

%-------------
   \begin{figure}
   \begin{center}
   \begin{tabular}{c}
   \includegraphics[height=10cm]{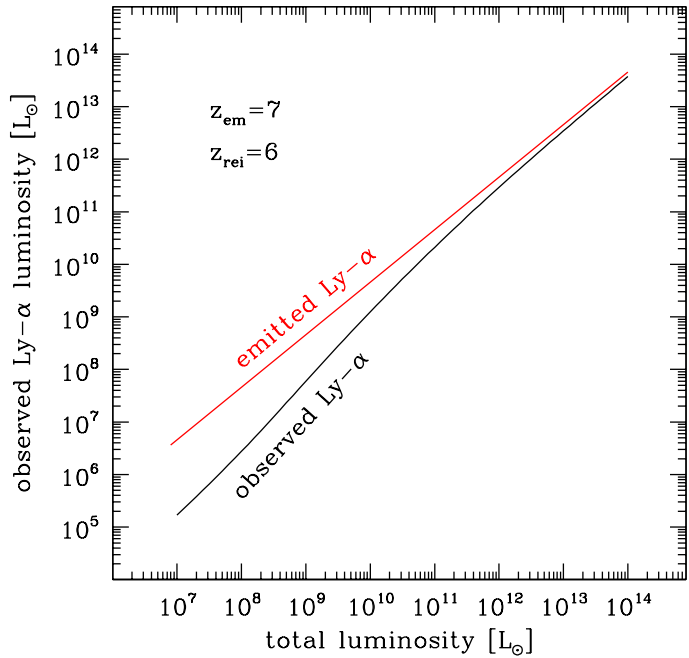}
   \end{tabular}
   \end{center}
   \caption[example] 
%>>>> use \label inside caption to get Fig. number with \ref{}
   { \label{fig:lymanalphaobserved} 
Transmitted Lyman $\alpha$ intensity as a function of the object
bolometric
luminosity. Brighter objects re-ionize their neighborhood and are
able to reduce the Lyman $\alpha$ attenuation.}
   \end{figure} 
%-------------

Another difference between zero-metallicity and low-metallicity HII
regions lies in the possibility that the latter may contain dust. If dust 
can form at low metallicity as well as in the local Universe, for
a $Z=10^{-3} Z_\odot$ HII region dust may absorb up to 30 \% of the
Lyman $\alpha$ line, resulting in roughly 15 \% of the energy being
emitted in the far IR\cite{Panagiaetal02}.

\section{FUTURE FACILITIES}
\label{sect:future}

It is natural to wonder whether primordial HII regions will be
observable with the generation of telescopes currently on the drawing
boards. In this section, I will focus mostly on the capabilities of the
Next Generation Space Telescope and of a 30m ground based
telescope. However, before considering these facilities in detail it
is necessary to estimate the effect of intergalactic absorption by
neutral hydrogen.

%-------------
   \begin{figure}
   \begin{center}
   \begin{tabular}{c}
   \includegraphics[height=10cm]{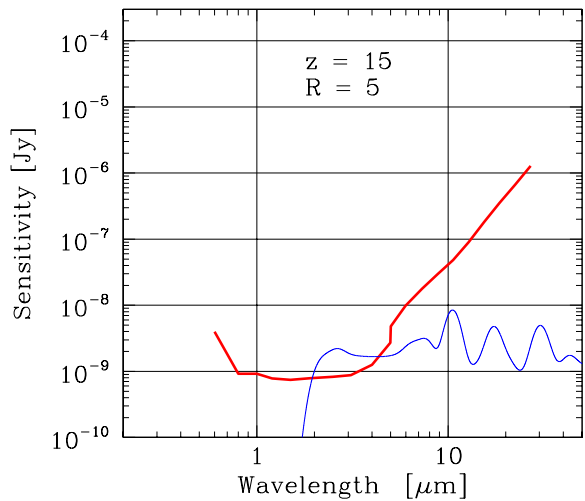}
   \end{tabular}
   \end{center}
   \caption[example] 
%>>>> use \label inside caption to get Fig. number with \ref{}
   { \label{fig:ngstimage} 
Synthetic spectral energy distribution of a Z=$10^{-3} Z_\odot$
starburst object at z=15 containing $10^6$ M$_\odot$ in massive stars
(thin line) compared to the imaging limit of NGST at R=5 (thick line).
The NGST sensitivity refers to 10$^5$s exposures with S/N=5. }
   \end{figure} 
%-------------
%-------------
   \begin{figure}
   \begin{center}
   \begin{tabular}{c}
   \includegraphics[height=10cm]{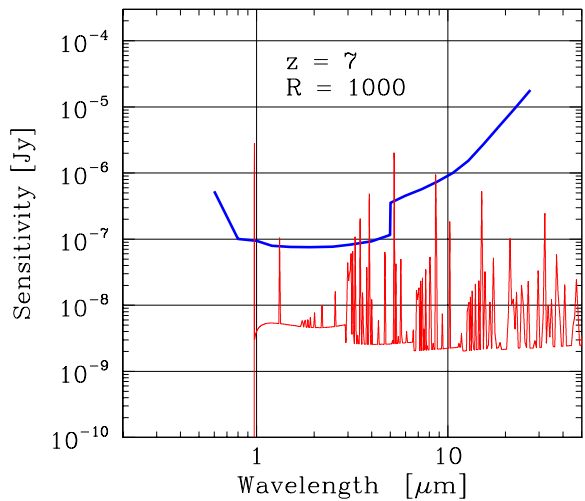}
   \end{tabular}
   \end{center}
   \caption[example] 
%>>>> use \label inside caption to get Fig. number with \ref{}
   { \label{fig:ngstspectra} 
Synthetic spectrum of a Z=$10^{-3} Z_\odot$ starburst object at z=7
containing $10^6$ M$_\odot$ in massive stars (thin line) compared to the
spectroscopic limit of NGST at R=1000 (thick line). The NGST sensitivity
refers to 10$^5$ s exposures with S/N=5. }
   \end{figure} 
%-------------

\subsection{The effect of intergalactic absorption}

Intergalactic absorption by intervening neutral hydrogen will suppress
Lyman $\alpha$ \cite{MiraldaEscude98,MadauRees2001,Panagiaetal02}.  A
comparison of the observed vs emitted Lyman $\alpha$ intensities is
given in Figure \ref{fig:lymanalphaobserved}.  The transmitted Lyman
$\alpha$ flux depends on the total luminosity of the source since this
determines the radius of the resulting Stroemgren sphere\cite{Panagiaetal02,Haiman02}. This effect may explain the possible detection of Lyman $\alpha$
at redshift greater\cite{Huetal} than that at which a Gunn-Peterson trough has been
detected in quasars\cite{Beckeretal2001}. A Lyman
$\alpha$ luminosity of $\sim10^9$ L$_\odot$ corresponds to $\sim10^6$
M$_\odot$ of massive stars. In the following I will consider a star
cluster of this luminosity.

\subsection{Imaging}

The synthetic spectra derived above can be convolved with a
suitable filter response in order to obtain a spectral energy distribution that
can be compared directly to the NGST imaging sensitivity for $10^5$s
exposures (an example is shown in Figure \ref{fig:ngstimage}). It 
is clear that NGST will
be able to easily detect such objects. Due to the high background from
the ground, NGST will remain superior to 30m ground-based telescopes
for these applications.

Scattering of Lyman $\alpha$ photons around a source before re-ionization
creates an extended ($\sim$15 arcsec at z=10) halo\cite{LoebRybicki} that is
in principle detectable in narrow band imaging by NGST. Unfortunately,
such a halo is over-resolved by NGST and would be detectable with S/N=10 
in a 10$^5$s exposure only around an object with a Lyman $\alpha$ luminosity of
10$^{12}$ L$_\odot$, {\it i.e.}, much brighter than the currently expected
population III sources.

\subsection{Spectroscopy}

The synthetic spectra can also be compared to the NGST spectroscopic
sensitivity for $10^5$s exposures (see Figure \ref{fig:ngstspectra}).
NGST is able to detect lines from a 10$^6$ M$_\odot$ starburst (in massive
stars)
only at relatively low redshift. It will
be possible to place interesting limits on the metallicity of these
sources only if they are either at a lower redshift or brighter than assumed
here.

Under somewhat special circumstances it may be possible to identify
objects at redshift higher the re-ionization also through the combined
signature of a Lyman $\alpha$ and a Lyman $\beta$
trough\cite{HaimanLoeb1999}. While potentially very interesting, this
particular observations would not confirm that these objects are truly
first generation.

Making spectroscopic observations at high spectral resolution (R$\gta$5000)
most of the
sky background in the H band (most relevant for z$\simeq$10-12) can be
reduced by a significant factor, so that the performance of a 30m
ground based telescope operating between the OH lines would
significantly exceed that of NGST. These telescopes would potentially
be able to help us identify true first generation objects by placing
strong upper limits on the intensity of metal lines.

\subsection{Conclusions}

It is possible to discern truly primordial populations from the next
generation of stars by measuring the metallicity of high-z star
forming objects. The very low background of NGST will enable us to
image first-light sources at very high redshifts, identifying them
through the Lyman break technique. Unfortunately, the relatively small
collecting area of a 6m NGST will limit its capability in obtaining spectra
of z$\sim$10 first-light sources. Thus, the discrimination between
first and second generation objects may have to rely on the next
generation of large ($\sim$30m) ground-based telescopes.

%%%%%%%%%%%%%%%%%%%%%%%%%%%%%%%%%%%%%%%%%%%%%%%%%%%%%%%%%%%%%
\acknowledgments     %>>>> equivalent to \section*{ACKNOWLEDGMENTS}       
 
Many results presented here have been derived in collaboration with Nino Panagia. I wish to acknowledge productive
discussion with N. Panagia and M. Fall.

%%%%%%%%%%%%%%%%%%%%%%%%%%%%%%%%%%%%%%%%%%%%%%%%%%%%%%%%%%%%%
%%%%% References %%%%%

\bibliography{report}   %>>>> bibliography data in report.bib
\bibliographystyle{spiebib}   %>>>> makes bibtex use spiebib.bst

\end{document}